\documentclass[10pt]{emulateapj}

\accepted{October 12, 2017}

\usepackage{times}

\usepackage{color}

\newcommand{\HII}{H$\;${\small\rm II}\relax}

\newcommand{\halpha}{H$\alpha$}

\newcommand{\msun}{M$_\odot$}
\newcommand{\kms}{km~s$^{-1}$\relax}

\begin{document}

\title{ALMA CO(3-2) Observations of Star-Forming Filaments in a Gas-Poor Dwarf Spheroidal Galaxy
}

\author{S. Michelle Consiglio\altaffilmark{1},
  Jean L. Turner\altaffilmark{1},
  Sara Beck\altaffilmark{2},
  David S. Meier\altaffilmark{3,4},
Sergiy Silich\altaffilmark{5},
  Jun-Hui Zhao\altaffilmark{6}}

\altaffiltext{1} {Department of Physics and Astronomy, University of
  California, Los Angeles, Los Angeles, CA 90095,
  smconsiglio@ucla.edu}

\altaffiltext{2} {School of Physics and Astronomy, 
  Tel Aviv University, Ramat Aviv, Israel}

\altaffiltext{3} {Department of Physics, New Mexico Institute of
Mining and Technology, Socorro, NM 87801}

\altaffiltext{4}{National Radio Astronomy Observatory, Soccorro, NM 87801 USA}

\altaffiltext{5}{Instituto Nacional de Astrof\'isica, \'Optica y Electr\'onica, Puebla, M\'exico C. P. 72840}

\altaffiltext{6}{Harvard-Smithsonian Center for Astrophysics, Cambridge, MA 02138 USA}

\begin{abstract}

We report ALMA observations of $^{12}$CO(3-2) and $^{13}$CO(3-2) in
the gas-poor dwarf  galaxy NGC 5253. 
These 0.3\arcsec(5.5 pc) resolution images 
reveal small, dense molecular gas clouds 
 that are located in
 kinematically distinct, extended filaments. Some of the
 filaments  
appear to be falling into the galaxy and may be fueling its current star formation.
The most
intense CO(3--2) emission comes from the central $\sim 100$ pc region centered
on the luminous  radio-infrared HII region known as the supernebula. 
The CO(3--2) clumps within the starburst region 
are anti-correlated with
H$\alpha$ on $\sim$5 pc scales, but are well-correlated
with radio free-free emission. 
Cloud D1, which enshrouds the supernebula, has a high
$^{12}$CO/$^{13}$CO ratio, as does another
cloud within the central 100 pc starburst region, possibly because
the clouds are hot. 
CO(3--2) emission alone does not allow determination of cloud masses  
as molecular gas temperature and column density are degenerate at
the observed brightness, unless combined with other  lines such as $^{13}$CO.

\end{abstract}


\section{Introduction}
\label{sec:intro}

Super star clusters (SSCs) are the birth places of most massive stars; these immense concentrations of
hot stars have the potential to  
have enormous impact on their host galaxies. Understanding the formation and evolution
of SSCs is critical to
understanding the energy budget, 
feedback, and regulation of star formation in galaxies and the
environments that produce them.

NGC 5253 is a local 
\citep[D=3.8 Mpc;][]{2004ApJ...608...42S} dwarf spheroidal galaxy undergoing 
an intense starburst that has created a large population of massive star clusters
over the course of a Gyr \citep{1995AJ....110.2665M,1996AJ....112.1886G,1997AJ....114.1834C,
2001ApJ...555..322T,2004AJ....127.1405C,2004ApJ...612..222A,2005A&A...429..449M,
2005A&A...433..447C,2013MNRAS.431.2917D,2015ApJ...811...75C,2016ApJ...823...38S}.    
NGC 5253 has a stellar mass of only $\sim 1.5 \times 10^8~$\msun\ \citep{1998ApJ...506..222M},
and a dark matter mass that is likely 8-9 times higher \citep{1999ApJ...513..142M}. 
This is a galaxy with dispersion-dominated kinematics and little rotation \citep{1982MNRAS.201..661A,1989ApJ...338..789C}. 
The abundant atomic hydrogen gas, comparable to the stellar mass at $\sim$2-3$\times$10$^{8}$
\msun\   
 \citep{1995ApJ...454L.121K,2008AJ....135..527K,2012MNRAS.419.1051L}, 
 resides largely outside the galaxy, in 
filaments and 
streamers extending to $\sim$3-5 kpc ($\sim$ 2-3 optical radii).

  The dominant radio/IR HII region in NGC 5253, the ``supernebula,'' is 
 excited by a massive, young SSC \citep{2000ApJ...532L.109T,2001ApJ...554L..29G,
  2002AJ....124..877M,2004ApJ...602L..85T}.
The supernebula has an 
ionizing rate of $N_{Lyc} = 3.3 \times
10^{52}~\rm s^{-1}$ \citep[][corrected for dust following
\citealt{2001AJ....122.1788I}]{2015Natur.519..331T,2017arXiv170706184B},
corresponding to an IR luminosity of $\sim 5\times 10^{8}\rm ~L_\odot$
\citep{2001ApJ...554L..29G},
contributing to a total galactic infrared luminosity of 1.6$\times10^{9}$ L$_\odot$  
 \citep{2005A&A...434..849H,2004A&A...415..509V}. 

  Many investigators find that there are two massive clusters in the nucleus
  \citep[e.g.][]{2004ApJ...612..222A,2015ApJ...811...75C},
  one associated with the supernebula and one with the \halpha\ peak;  however, 
 extinction within the central starburst region is very high, making it
  difficult to match radio, infrared, and optical observations
\citep{1997AJ....114.1834C,2003Natur.423..621T,2005A&A...429..449M}. 
Abundant molecular gas associated with such a luminous HII region
  in the ultracompact stage is expected, but the CO(1-0) and (2-1) emission is weak. 
Roughly three-quarters of this CO emission, as well as the HI, is found along the minor axis,
outside the central regions
\citep{1997ApJ...474L..11T,2002AJ....124..877M,2015PASJ...67L...1M}. 

Previous observations of CO(3--2) 
from the Submillimeter
Array (SMA)  
indicate that the star formation in the CO cloud (Cloud D) near the
supernebula is very efficient, having converted 
$>$50\% of its gas mass into stars \citep{2015Natur.519..331T}. 
The unexpectedly bright CO(3--2)
emission in the central starburst region, 
as compared to lower J CO lines, indicates that this central cloud, Cloud
D \citep{2002AJ....124..877M}, is  warm, $T\sim 300$ K. 
However, the $\sim$4\arcsec $\times$2\arcsec\ beam of the SMA images 
does not separate the dense gas at the 
cluster scale. 

With
higher resolution and sensitivity, the Atacama Large Millimeter/Submillimeter Array (ALMA) can 
image NGC 5253  at the $\sim$5 pc cluster scale. We have been able to image the structure
of its remarkable starburst. 
Here we
  present 2015 ALMA observations of the J=3--2 rotational transition of CO
  in NGC 5253  with 0.3\arcsec\ ($\sim$5.5 pc) resolution.

\section{Observations}
\label{sec:obs}
NGC~5253  was observed in ALMA Band 7 
as a Cycle 2 (Early Science)
program (ID = 2012.1.00105.S, PI = J. Turner) 
on 2015 4 and 5 June. 
 Two fields, each with an 18\arcsec\ field-of-view, centered on
13:39:56.62 -31.38.33.5 and 13:39:55.91 -31.38.26.5
(ICRS) 
were observed simultaneously with 8383 seconds on both sources. 
The uv range covers
$\sim$24.5-900 k$\lambda$;
the largest structures sampled are 4-8\arcsec.
Spectral resolution of 244.14
kHz, or 1 \kms\ per channel resolves the CO(3--2) lines.
Bandpass, flux, and phase  were calibrated
with J1427-42064, Titan,  and J1342-2900, respectively. Absolute flux
density calibration is estimated to
be better than 10\% in Band 7 Cycle 2 \citep{Lundgren_2013}. 
Data calibration was performed using the code produced by the Joint ALMA Observatories 
 with CASA (Common Astronomy Software Applications). 
Continuum emission was subtracted in the (u,v) plane before creating line maps.
The beam size is 0.33\arcsec\ x 0.27\arcsec\  at  p.a.=-89.8$^o$ for
$^{12}$CO(3--2) and 0.33\arcsec\ x 0.29\arcsec\ at p.a.=89.0$^o$ for $^{13}$CO(3--2).
The conversion to brightness is such that 1 K corresponds to $\sim$10-17 mJy, depending
on source geometry. 
The RMS in a single 1 \kms\  channel  is 2.7~mJy/bm for 
$^{12}$CO(3-2) and 3.2~mJy/bm for $^{13}$CO. Integrated intensity images (MOM0) were created by summing emission
in all channels that was greater than $\pm 1\sigma$; for the intensity-weighted mean velocity
map (MOM1); the cutoff was  $\pm 5\sigma$. 

The single dish integrated flux density reported by \citet{2001AJ....121..740M} with the 22\arcsec\ CSO beam was
3.4 K km/s, corresponding to $\sim$155--170 Jy $\rm km~s^{-1}$,
depending on assumed source geometry.  
The integrated flux density recovered in the lower resolution
SMA synthesis map of \citet{2015Natur.519..331T}
was 110 $\pm 20$ Jy $\rm km~s^{-1}$.
We detect $\approx$50\% of the expected integrated flux density the SMA maps in
the form of individual clumps;
  the remaining flux density is extended and much has
  been resolved out because of the lack of short spacings in the array. 
 Although the individual dense clouds that emit CO(3-2) are not likely
 to be large, collections of such clouds with low  ($\lesssim$10\%)
 volume filling may lie below 
 even the ALMA sensitivity limit.

\section{The $^{12}$CO(3--2) Line: How is the Dense Gas in NGC 5253
  Distributed?}
\label{sec:co32}
\subsection{The Dense Gas is Clumpy and the Clumps Form Filaments}

\begin{figure}
\includegraphics*[width=\linewidth]{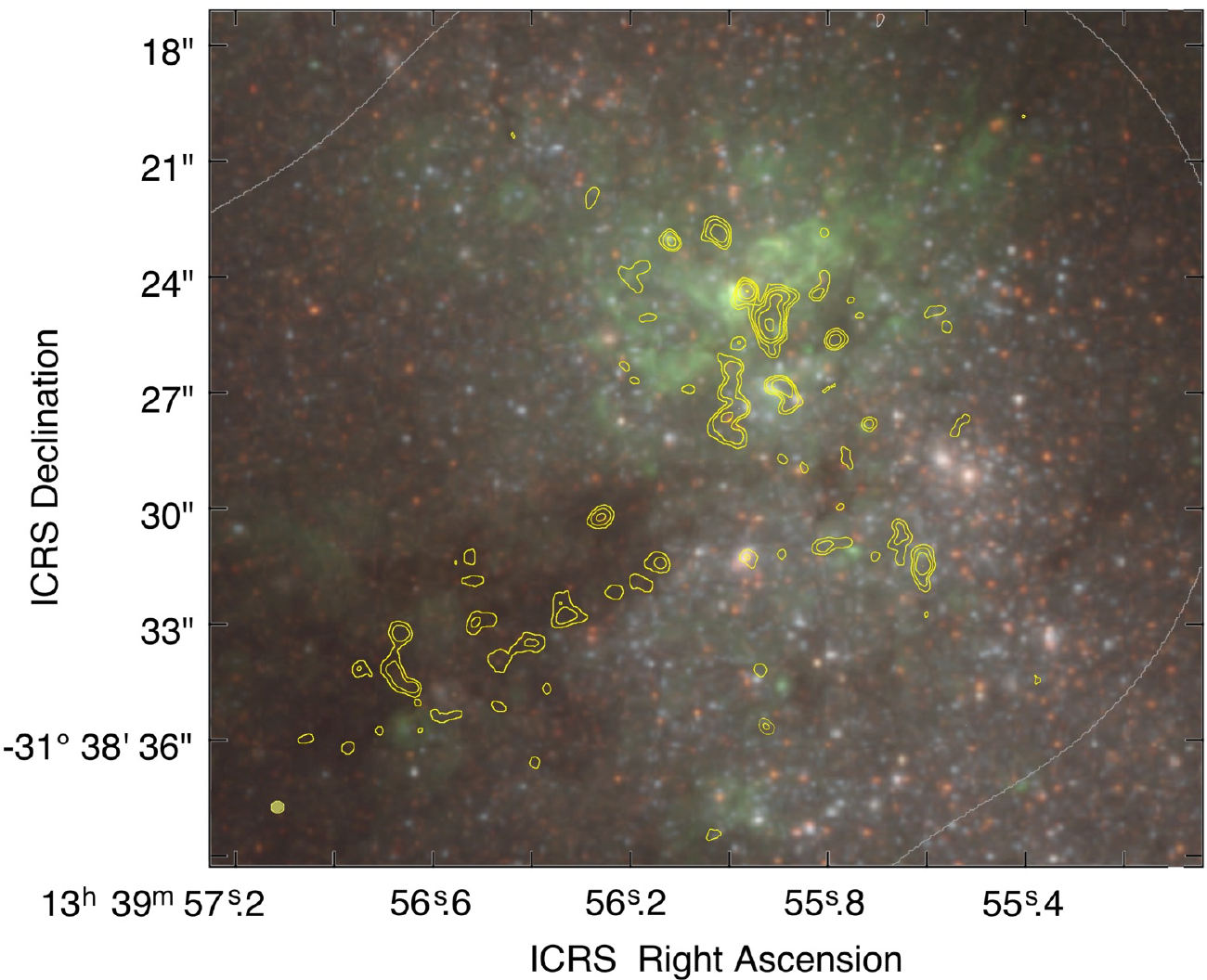}
\caption{ALMA CO(3-2) contours   are overlaid on 
an archival HST/ACS image of NGC~5253 in the 814W, 555W, 435W bands in
red, green, and blue channels respectively, showing the stars. 
Contours are $2^{n} \times
0.1$~Jy~bm$^{-1}$~km~s$^{-1}$. CO(3--2) falls along the dust lanes. Registration
is estimated to be better than 0.3\arcsec, determined by aligning the supernebula/Cloud D1 with 
the infrared nebula (Turner et al.\ 2017; D. Cohen et al., in prep.). The half 
power of the primary beams of the two mosaicked pointings (field of view) is in outline at the edges
of the figure. ALMA beam is 0.33\arcsec $\times$ 0.27\arcsec,
p.a.=-89.8\degr, shown at bottom left.}
\label{fig:HST_CO}
\end{figure}

The 
CO(3-2) integrated line intensity, S$_{CO}$=$\int{I_{CO}dv}$,  in 
NGC~5253 
 is shown in Figure~\ref{fig:HST_CO}, superimposed on an archival 
HST/ACS image of the galaxy.   The J=3 state of CO
has an excitation energy of
33K, requiring a critical density of $n\gtrsim 20,000~\rm cm^{-3}$ for
excitation, thus the J=3 state of CO 
tends to trace warm and dense star-forming
molecular clouds
\citep{2010ApJ...724..687L,2010ApJ...723.1019H}. 

\begin{figure}
\includegraphics*[width=\linewidth]{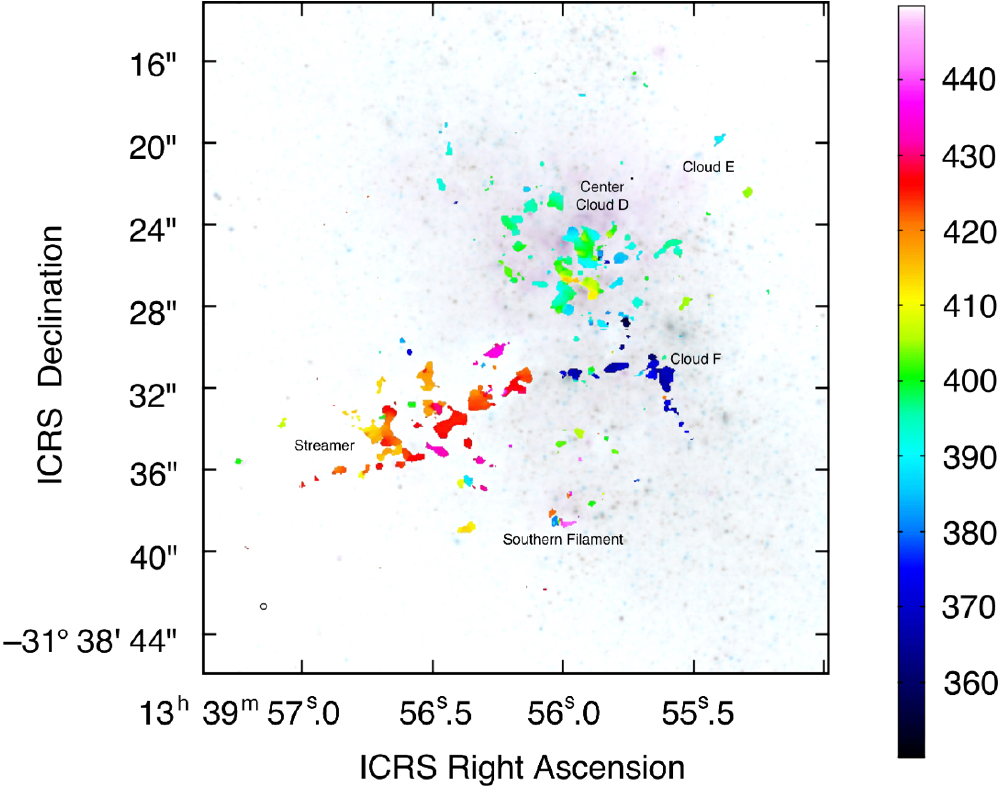}
\caption{Intensity-weighted mean velocity, ``MOM1", image of CO(3--2) in NGC 5253.  
  Regions and filaments discussed in the text are labeled; the velocity scale, barycentric, is shown at right. 
  The MOM1 map is constructed from 
emission $>$5$\sigma$ in absolute value. It is shown superposed on an HST F555W image. The 
figure shows that the 
filamentary structures that appear in the integrated intensity map of Figure 1 are also
distinct kinematically. The ALMA beam is shown at the bottom left.}
\label{fig:Mom1}
\end{figure}

Within the $\sim$ 30\arcsec\ by 60\arcsec\ region covered by the two ALMA pointings,
  CO is detected over an area $\approx$20\arcsec or 370 pc.
We identify five regions, labeled in Figure \ref{fig:Mom1}, whose properties are given in
Table \ref{tab:13co}. 
In addition to 
``Cloud D'' \citep{2002AJ....124..877M, 2015Natur.519..331T} which has the brightest CO(3--2) clumps, 
there is
 the ``Streamer,'' which lies along the optical dust lane to the
 southeast, ``Cloud E,'' as observed in \citet{2002AJ....124..877M}, northwest of the center, 
 ``Cloud F'' from \citep{2015Natur.519..331T} to the southwest, and
 finally, the ``Southern filament,'' 
 a group of clumps due south of the center.
Individual filaments have characteristic widths of only
a few clumps, $\sim$0.5\arcsec, but typical lengths of  
$\sim$5\arcsec or 90 pc. 
Individual clumps at these densities have gravitational collapse free-fall timescales of t$_{ff}\lesssim$3.6$\times$10$^{5}$~yr.

Previous observations of NGC 5253 
established that lower J CO emission is located primarily in the Streamer
\citep{1997ApJ...474L..11T, 2002AJ....124..877M,2015PASJ...67L...1M,2015Natur.519..331T}, following
the prominent dust lane to the east along the minor axis of the galaxy. 
Optical emission
lines, probably from the surface of the streamer, are also present in this feature 
\citep{1981PASP...93..552G,1998ApJ...506..222M,2011ApJ...741L..17Z}. 
 At the 10 times higher spatial resolution of the ALMA images 
the Streamer  splits up into clumps
roughly a beamsize ($\sim$0.3\arcsec, 5 pc) in extent or less, arranged in filamentary structures
that are $\sim$ 25--200 pc in extent.

The properties and kinematics of the clumps within the filaments are discussed in this section.
 Individual clouds in the central starburst region, within Cloud D, are discussed further 
in $\S$5.

\subsection{The Clumps Form Filamentary Structures that are
  Distinct in Velocity} 
Kinematic information is important in interpreting the molecular gas
structure in NGC 5253. The systemic velocity of the galaxy is not well
defined. HI observations show a systemic velocity of $\sim$ 400 \kms
\citep{2004AJ....128...16K}. 
However, the HI lies in the outskirts of the galaxy, is asymmetric,
and is affected by large infalling streamers 
\citep{1995ApJ...454L.121K,2008AJ....135..527K,2012MNRAS.419.1051L}. 
The H$\alpha$ images of \citet{1982MNRAS.201..661A} 
 show a central velocity of $\sim$385 \kms. 
Since H$\alpha$ is more closely associated with the stellar
distribution, this is probably closer to the systemic velocity. Our ALMA
CO(3--2) intensity-weighted mean velocity (MOM1) image is shown in
Figure 2. This image reveals that the median CO radial (radio
definition, barycentric) velocity within the central 100 pc is
$\sim$380 \kms. We adopt 380 \kms~ as the central velocity,
although it is not clear how our v$_{sys}$ relates to the stellar velocity.

The filamentary structures identified in the integrated intensity
image 
are not only spatially but kinematically distinct.  
The Streamer consistently runs 
$\sim$~40--20~\kms~redward of the systemic 
velocity v$_{sys}\sim 380$~\kms. 
Cloud F is $\sim$~20--30~\kms~ blueward,  and the Southern filament is
$\sim50$~\kms~ to the red.  Clouds D, in the central region, and  E are closer to 
the systemic velocity. 
Thus, each filament
is distinctly defined kinematically and spatially, formed of
gas that is sufficiently dense to form stars. The HI cloud surrounding
NGC 5253 is also composed of filaments. The CO filaments of the
Streamer have been identified as forming the inner portion of   HI
filaments \citet{2008AJ....135..527K}. 
The other CO(3--2) filaments may also form the inner portion of more extended HI
filaments.

Based on radial velocity 
alone it cannot be determined  
if red- or blue-shifted emission is inflowing or outflowing. However, 
many of the filaments correspond to areas of extinction in visible
images, indicating that they are located on the near side of the galaxy. 
Since both the Streamer and the Southern filament are redshifted and foreground,
they are thus falling toward NGC 5253. On this basis, it has previously been proposed that
the starburst in NGC 5253 has been fueled by accretion from this dust lane
 \citep{1997ApJ...474L..11T,2002AJ....124..877M}.

 By contrast Cloud F to the southwest of the central starburst is blueshifted. Cloud F 
does not lie in an area of obvious extinction. Thus, we propose that Cloud F represents an infalling filament
on the far side of the galaxy.

There is no clear velocity structure within individual filaments, which seem to have
coherent velocities. In particular, we find no evidence of acceleration in the gas along the Streamer,
as would be expected if it were in free-fall towards the galaxy. Simple models of gravitational free-fall,
assuming a 
  centrally-concentrated mass structure, 
  predict
  acceleration of the gas towards the center that is not clearly
  visible in
  the data shown in Figure~\ref{fig:Mom1}.
The lack of
evidence for acceleration was also noted by
\citet{2008AJ....135..527K} in the associated HI filament. 
We note, however, that  the
gas could be infalling nonetheless. Gas experiencing drag can fall
towards a central mass at constant velocity. 

\subsection{Conditions in the Filament Clumps}

The  clumps typically have diameters of about a beam, $\la$5 pc,  
with $^{12}$CO(3--2) integrated line intensities of 0.1-0.5 Jy $\rm km~s^{-1}$.  
Clumps have been individually identified
using cprops,a clump finding code designed to identify clumps using velocity and spatial information
\citep{2006PASP..118..590R}. Their individual characteristics will be published
elsewhere 
\citep{ConsiglioXXX2017}. Masses for the larger clumps and features are 
listed in Table 1. 
To obtain clump masses from the CO emission, we use the CO conversion factor,
$X_{CO}$, which relates CO(1--0) integrated line intensities to H$_2$ gas masses.
CO(1--0) is very weak  in NGC~5253
  \citep{1997ApJ...474L..11T}.  Thus, CO(1--0) integrated line
intensities are estimated
from the CO(3--2) using the line ratio, 
R$_{31}$=$\frac{I_{CO32}}{I_{CO10}}$=7, observed in
II~Zw~40  \citep{2016ApJ...833L...6C}, a similar low-metallicity star-forming
galaxy.  
The II~Zw~40 value is appropriate since  the deep ALMA CO(3--2) and CO(1--0) 
images used for that calculation 
  had matched beams of $\sim$0.3\arcsec, similar to the beam here, in
  a galaxy with a similar enrichment and metallicity environment. 
This ratio corresponds to optically thick, slightly subthermal
gas.
 However, the unknown value of R$_{31}$ in NGC 5253 is a source of systematic uncertainty
  in the gas masses. 
  For a conversion factor of X$_{CO} = 4.7 \times 10^{20}$
cm$^{-2}$ (K km$^{-1}$)$^{-1}$, the CO(3--2) integrated line
intensities correspond to clump masses
of 3.3-35$\times$10$^{4}$~$\rm M_\odot$ (the largest clumps are listed in Table 1). 
The conversion factor mass may overestimate the
masses in Clouds D1 and D2, which are likely to 
contain significant stellar mass 
\citep[see the discussion in $\S$\ref{sec:13co}, and ][]{2017XXXT}. 
These individual 
dense clumps by themselves are not large enough to form the large, $\sim 10^{4}-10^{6}$~\msun,  clusters that are
currently seen in NGC 5253 \citep{2013MNRAS.431.2917D,2015ApJ...811...75C}.

We estimate that 
individual clouds in the filaments have internal pressures $P/k=nT \gtrsim 2-4\times 10^{5}$ cm$^{-3}$K
for temperatures of $\sim$20K, and higher for Cloud D, which appears to be warmer. 
This value  is higher than typical 
ambient pressures in the interstellar medium, 
P/k=2250 cm$^{-3}$K \citep{2003ApJ...587..278W}, of our Galaxy.

\section{Dense Gas and Star Formation in NGC 5253}
\label{sec:starformation}

The central $\sim$150 pc region surrounding the supernebula in NGC
5253 
\citep[``Cloud D" of][]{2002AJ....124..877M} contains an 
extraordinary
concentration of young stars. 
Here we use ALMA CO(3-2) images and previous lower resolution observations
of CO(2--1) and CO(3--2) to understand the properties of dense gas and
related star formation. 

\begin{figure}
\includegraphics*[width=\linewidth]{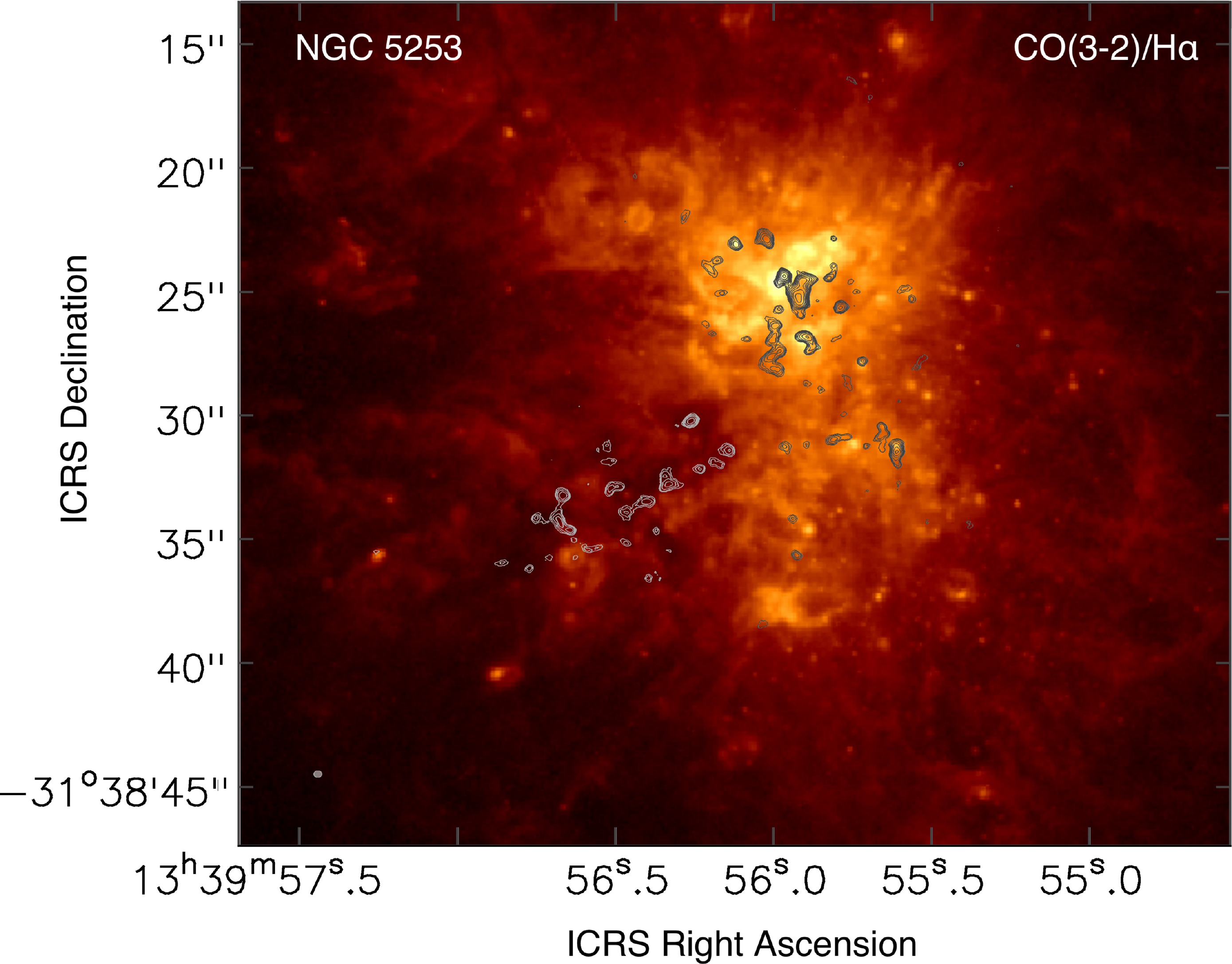}
\caption{ALMA CO(3-2) image and H$\alpha$. The CO(3-2) contours overlaid on an 
archival HST image of H$\alpha$ emission in color. Contours are $2^{n/2}$ times
80 mJy~bm$^{-1}$~km~s$^{-1}$.
The CO(3-2) clumps are 
generally anti-correlated
with the H$\alpha$, consistent with the $A_V \ga 4$ expected for the
clumps. 
ALMA beam, 0.33\arcsec $\times$ 0.27\arcsec,
p.a.=-89.8\degr, is shown at bottom left.}
\label{fig:Halpha_CO}
\end{figure}

\begin{figure}
\includegraphics*[width=\linewidth]{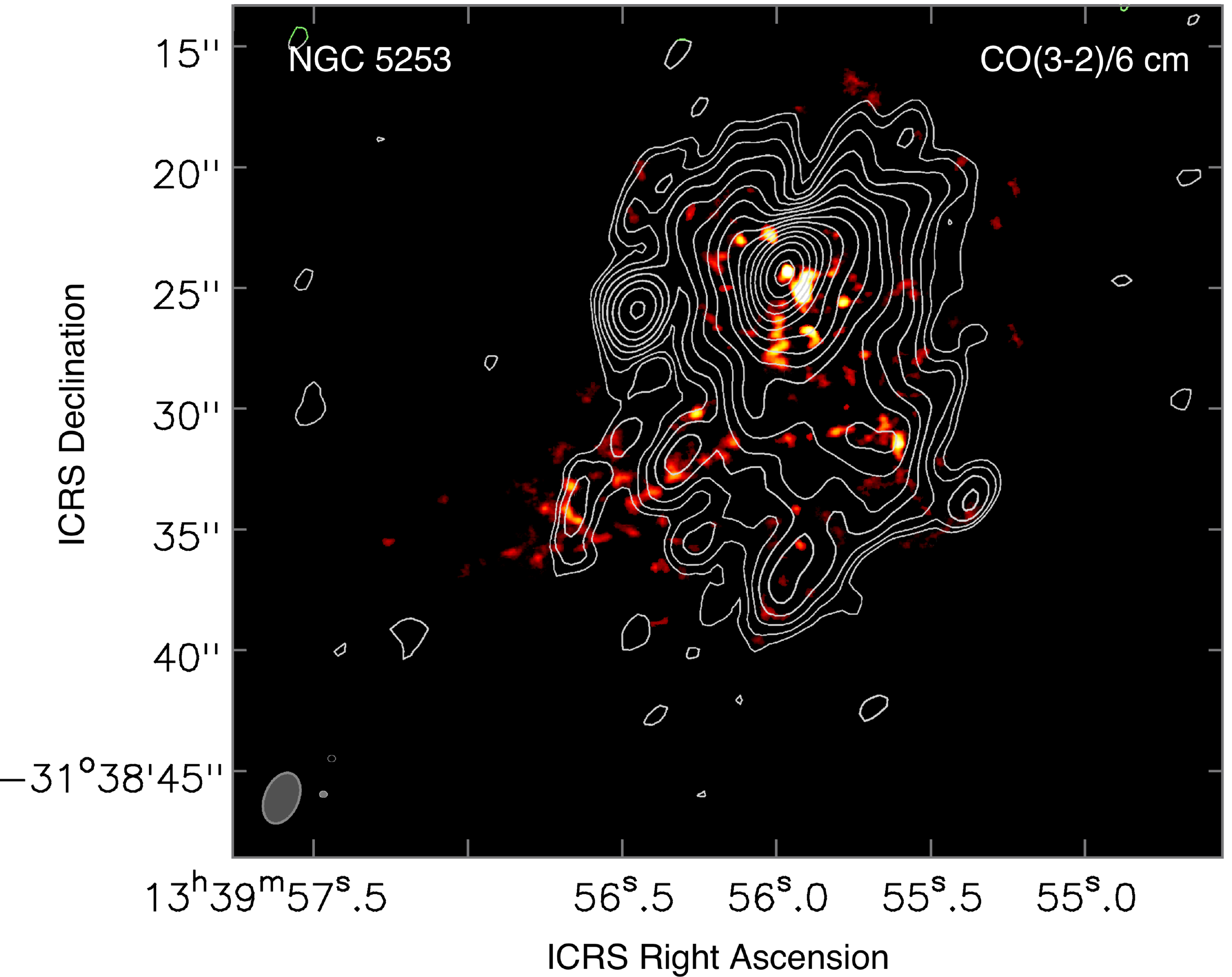}
\caption{ALMA CO(3-2) image and 6cm continuum emission. 
The CO(3-2) shown in color, over the integrated intensity range of
$\sim0.015-4.2$~Jy/bm~km/s. 
The VLA 6 cm emission is contoured at levels of $\pm0.07 \times 2^{n/2}$~mJy/bm (from Turner et al.\ 1998).
 The VLA beam is $\sim$2\arcsec, which
is much larger than the ALMA CO(3-2) beam of $\sim$0\farcs3; both beams are shown at the lower left. 
Nearly all the 6~cm emission is thermal free-free emission from HII regions, which in the absence
of extinction, will look like H$\alpha$. The exception is that the 
strong continuum point source at R.A.$\approx 56.5^s$ to the east of the central source is a supernova
remnant (Turner et al. 1998).}
\label{fig:Halpha_6cm}
\end{figure}

The CO(3--2) integrated line intensity is shown with
two tracers of star formation, H$\alpha$ and radio continuum,
in Figures~\ref{fig:Halpha_CO} and \ref{fig:Halpha_6cm}. 
 The 
CO(3--2) clumps in the ALMA image and the  \halpha\ emission
in NGC 5253 do not show an obvious correlation (Figure~\ref{fig:Halpha_CO}). In
fact on $\sim$5 pc scales these properties appear to be  anti-correlated, with CO(3--2) appearing in dust
lanes in the H$\alpha$ image. The correspondance of CO with dust lanes is not
surprising: CO forms at $A_V \gtrsim 2$ \citep[e.g][]{2015ApJ...803...37B},  
corresponding to an  overall cloud column of 
$A_V\gtrsim 4$. 
 The bright central HII region for which NGC 5253
is so well known \citep{1962ApJ...135..694B} 
is centered on the heavily extincted supernebula \citep{2003Natur.423..621T,2004ApJ...612..222A,
2004ApJ...602L..85T}. 
Loops and arcs
in the H$\alpha$ suggest that the supernebula is responsible for not only large scale
leakage of photons, 
but also possibly winds. 
Both effects
suggest patchiness of extinction in the supernebula core, consistent with the
kinematics and brightness of Cloud D1 \citep{2017XXXT}.

Radio continuum emission traces ionizing photons through free-free emission;
like \halpha, it traces emission measure, but is far weaker than \halpha. 
Radio continuum is the most reliable tracer of star formation in regions of high extinction. 
In Figure~\ref{fig:Halpha_6cm},  a VLA 6 cm image  
\citep{1998AJ....116.1212T} is overlaid on
 the ALMA CO(3--2) map. 
Except for an isolated supernova remnant 
to the east, the remaining 6 cm emission is  thermal free-free emission from \HII~ regions. 
The $\sim$ 2\arcsec\ resolution of this VLA image is much lower than
the ALMA images, but 
an excellent spatial correlation of the free-free continuum
with CO(3--2) is
apparent.  
The dominant radio continuum source in this image is the supernebula, 
associated with Cloud D1. Radio continuum peaks are also 
associated with Clouds F
and the Southern Filament, located $\sim 140-180$ pc to the south and
west of the supernebula, as well as clouds to the northwest within the Cloud D supernebula
complex. The Streamer also 
shows radio continuum emission, presumably from the same gas producing the 
``ionization cone" seen in [OIII] $\lambda$5007, H$\alpha$, [SII] and [SIII] 
$\lambda\lambda$6716, 9069 
\citep{1981PASP...93..552G,1998ApJ...506..222M,2011ApJ...741L..17Z}. 

The relation of gas, star formation, and molecular gas
depletion time can be estimated from the radio continuum and CO(3--2). 
Because  free-free emission is weak and difficult to
  detect at subarcsecond scales, 
we quantify the relation of the 6 cm free-free and CO using the VLA maps
 with larger 1-2\arcsec\ (20-40 pc) beams \citep{1998AJ....116.1212T}
  and images of  CO(2--1) from the Owens Valley Millimeter (OVRO)
Array  \citep{2002AJ....124..877M} 
for gas masses.  
The Streamer has the brightest emission in low J CO images, and contains most of the 
molecular mass in NGC 5253. 
The estimated gas mass of the streamer based on CO(2-1)
is $\sim 1.7\times 10^6$ \msun\  
 \citep[corresponding to Cloud ``C" of][corrected to $X_{CO}=4.7 \times 10^{20}\rm
 ~cm^{-2}\,(K ~km^{-1})^{-1}$]{2002AJ....124..877M}. 
The gas surface density based on their 9.6\arcsec\ $\times$ 4.8\arcsec\ beam is
$\sim$ 90~\msun~pc$^{-2}$, 
comparable to surface densities inferred for 
GMCs in the Galaxy \citep{2010A&A...519L...7L}.
Using Starburst99 models and a full Kroupa IMF, the relation between young stellar
mass and Lyman continuum rate is $\log [(M/M_\odot)/(N_{Lyc}/s^{-1})] = -46.6$,
with $N_{Lyc}=1.25\times10^{50} 
{\rm s^{-1}} (T/10^{4} K)^{-0.507} 
(\nu/{\rm 100 GHz)}^{0.118}
 D_{Mpc}^{2}S_{3.3mm} 
\rm(mJy)$.
For a mean star formation timescale 
of 5 Myr for the star formation, and for a radio continuum flux
of $\sim$1 mJy for  Cloud C \citep{2002AJ....124..877M}, 
which contains most of the Streamer emission, the estimated mass depletion
timescale is $\tau_{dep}$=M$_{gas}$/SFR$\approx$0.3 Gyr for the Streamer.
We estimate that this number is uncertain to factors of 3--5, due to the unknown star
formation timescale and uncertain gas mass.

The central 1\arcsec\ ($\sim 18.4$~pc) region surrounding the supernebula, Cloud D, is forming stars at
a rate of SFR$\sim$0.1 M$_{\odot}$ yr$^{-1}$, \citep{2015Natur.519..331T, 2017arXiv170706184B}.
CO(2--1) images indicate that M$_{H_{2}}\sim$8$\times$10$^{5}$ M$_{\odot}$ within
Cloud D. If this gas is all associated with the current star
formation, the gas depletion timescale is $\tau_{SF}\sim$8 Myr. However, these ALMA
maps reveal that Cloud D actually consists of many clumps ($\S$5), and it is unclear
if all of them are related to the currently forming massive cluster. If some of this
gas is not associated with the current star formation as suggested by the gas kinematics
\cite{2017XXXT},  then this is an upper limit to
$\tau_{SF}$.

The molecular gas depletion time in the Streamer, $\tau_{dep}= 0.3$~Gyr, is  
 less than the median value of 0.8 Gyr 
for local low metallicity galaxies 
as determined by \citet{2011AJ....142...37S}, but  within the observed spread.  
While CO-dark H$_2$ may be partly responsible for the low value, presumably the same holds for
the general low-metallicity sample as a
whole. It may also be that the ionized 
gas of the Streamer is caused by uv photons from the nuclear starburst, and not from
star formation within the streamer, which would increase the value of $\tau_{dep}$.

Cloud D has a very short molecular gas depletion time of $\tau_{dep}= 8$~Myr, a factor
of $\sim$100 times smaller than the typical low metallicity galaxy in the
\citet{2011AJ....142...37S} sample.
We caution that the central gas mass is uncertain, and the star formation
  rate depends on our assumption of 5 Myr for the age; the latter is probably uncertain to
  factor two, although there is an additional uncertainty in the spread in star formation age.
  However, these uncertainties of a factor of a few are not expected to be large enough
to explain the two order of magnitude shorter depletion time in Cloud D1. This star formation
is unusually efficient.

\citet{2013ApJ...778..133L}
point out that individual GMCs in the Galaxy do not obey a Kennicutt-Schmidt law; the star formation
rate surface density varies more than two orders of magnitude among
Galactic GMCs. They propose that the 
Schmidt-Kennicutt law for galaxies \citep{1989ApJ...344..685K,1998ApJ...498..541K}
is  a function of
relative dense gas fraction and beam dilution on kpc scales. 
With these high resolution ALMA
images 
we resolve individual GMCs in NGC~5253, 
and thus it is not surprising that we see a 
spread in star formation surface densities similar to that observed in Galactic GMCs. However by Galactic
standards, Cloud D is 
extraordinarily efficient at forming stars.

\section{Analysis of $^{13}$CO in the Clumps within Cloud D}
\label{sec:13co}

\begin{figure*}
\includegraphics*[width=\linewidth]{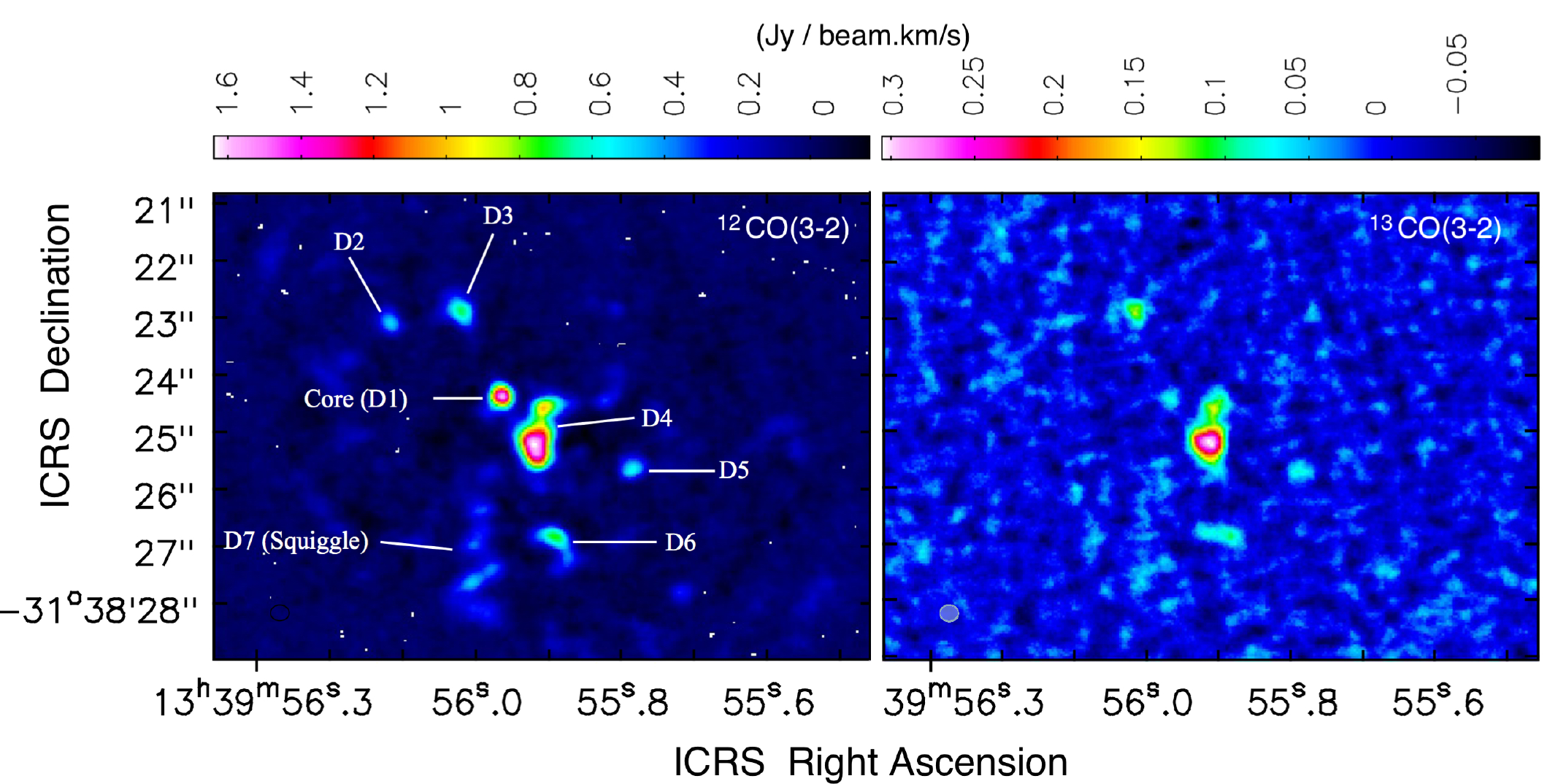}
\caption{The Central Region (Cloud D).  {\it Left)} ALMA 0.33\arcsec\ x 0.27\arcsec\
  $^{12}$CO(3--2) map with central regions labeled. RMS in a 1 km/s channel is 2.7 mJy/bm. 
{\it Right)}  ALMA 0.22\arcsec\ x 0.29\arcsec\ $^{13}$CO(3--2)
map. RMS in a 1 km/s channel is 3.3 mJy/bm.}
\label{fig:center}
\vspace{0.1in}
\end{figure*}

Images of these $^{12}$CO(3--2) and $^{13}$CO(3--2) 
for the central $\sim 100$~pc region corresponding to Cloud D 
are shown in Figure
\ref{fig:center}. With the high, $\sim$5~pc,
resolution of the ALMA images, the central Cloud D detected 
in CO(3--2) images from the Submillimeter Array \citep{2015Natur.519..331T}, 
breaks into a number of smaller clouds with diameters of $\sim 5-20$~pc.

The  larger CO structures 
are labeled and their
characteristics 
are detailed in
Table \ref{tab:13co}.
$^{13}$CO(3--2) emission provides valuable additional information
on the optical depth of the CO and the gas mass in the central
star-forming region of NGC 5253.
All clouds  in the central region are at least weakly detected in
$^{13}$CO. 
 $^{13}$CO(3--2) emission is also
detected for the brighter clouds in the Streamer, not shown here; but
this emission is very faint $\lesssim$2$\sigma$.

Clouds D3, D4, D5, D6, and Cloud F have $^{12}$CO(3-2)/$^{13}$CO(3-2)
flux ratios of $\sim 6-8$. 
These values are somewhat lower, than the value 
 of $\sim$13 observed in Orion \citep{1997ApJS..108..301S}.  
We infer that, as in Orion, these clouds
are optically thick in $^{12}$CO(3-2). 
In fact, 
 half of the measured 2.4 Jy km s$^{-1}$
of $^{13}$CO(3-2)
emission arises in 
Cloud D4, an apparently massive cloud in the
central region located about 15-20 pc south and west of the supernebula that 
does not host obvious 
star formation and which 
has  an optically thick $^{12}$CO/$^{13}$CO ratio of 7. 
Cloud D4 also appears in H$_2$ $\lambda$2.12$\mu$m images of the
region \citep{2005A&A...433..447C}, suggesting the presence of shocks (turbulence)
or fluorescence; given the extinction implied by the CO, it is likely that the H$_2$ 
emission arises on the near surface of the cloud. 

The highest  $^{12}$CO(3--2)/$^{13}$CO(3--2) ratio is $\sim$40, for Cloud D1, 
 the small (r$\sim$2.8 pc) cloud coincident with the giant central
HII region known as the 
supernebula. This value is uncertain because the $^{13}$CO(3--2) is
so weak, but is unlikely to be $<$20. Star formation may still  be
ongoing within D1, which appears to coexist
with the  $\sim 2.5 \times 10^5$~M$_\odot$ super star cluster and HII region
\citep{2017XXXT}. 
Based on the high   
$^{12}$CO(3--2)/$^{12}$CO(2--1) ratios of lower resolution maps, 
  \citet{2015Natur.519..331T} suggested that the larger 
Cloud D is optically thin.    The $^{12}$CO(3--2)/$^{13}$CO(3--2) ratio
for D1 is consistent
with the suggestion.
This situation would be unusual; 
 Galactic GMCs are  optically thick in these lines. 
Cloud D1 
may be 
optically thin because the gas is hot \citep{2015Natur.519..331T} and the molecules
spread out over the rotational ladder.  
At 300K, the partition 
function is 100, such that only 1\% of the CO molecules will be in the ground state
and $\sim$3\% in J=1. 
It is also possible that Cloud D1 could  also have selective
photodissociation or that the $^{12}$CO(3--2) and $^{13}$CO(3--2) could probe 
different physical environments due to temperature gradients in cloud
surfaces due to the presence of strong radiation fields in the cluster environment. 
Cloud D1 is
also bright in  H$_2~\lambda 2.12~\mu$m emission \citep{2005A&A...433..447C}; 
given that Cloud D1 is likely to consist
of multiple star-forming clumps, the H$_2$ emission may arise on the surfaces of these 
individual clumps. (However Clouds D4 and D6, which are  optically thick, are also bright
in the H$_2$ line.)

Cloud D2, located $\sim$40 pc
  to the northeast of Cloud D1, 
  may also be optically thin, with a
  $^{12}$CO/$^{13}$CO ratio of $\sim$12, 
 which  is higher than average for
  this galaxy. As this cloud appears in radio continuum 
  emission at cm wavelengths (Beck+18, private
  communication) and  in H$_2~\lambda 2.12~\mu$m emission 
\citep{2005A&A...433..447C}, 
Cloud D2 appears similar to Cloud D1, if smaller in scale.
  Thus, Cloud D2 appears to be a young, compact cluster that has warm 
  molecular gas, although significantly smaller than the 
  massive cluster within D1.

If $^{13}$CO(3-2) is optically thin in some clouds, we can use the $^{13}$CO 
integrated line intensities to estimate optically thin masses. These masses are given in 
Table 1. 
For Cloud D1 we adopt a temperature of 300 K, based on RADEX
models of the CO(3--2)/CO(2--1) line ratio \citep{2015Natur.519..331T}
and for clouds besides D1 we adopt 20K. The optically thin
masses are not very sensitive to temperature for this line until temperatures exceed $\sim$100K,
although if the clouds are actually very warm, these masses will be underestimates. 
If we adopt a Galactic [CO]/[H$_{2}$] ratio of $8.5\times 10^{-5}~$, and a
 [$^{12}$CO]/[$^{13}$CO] abundance ratio of 40, the minimum indicated by the Cloud D1 line ratio,
  we obtain cloud masses M$_{Cl}\sim$0.6-11$\times 10^{3} M_\odot$ (Table 1), 
 with the most massive
 cloud being the apparently quiescent 
 Cloud D4). 
 
 The masses based on $^{12}$CO(3-2), assuming 
a conversion factor (``$^{12}$CO Masses") and those 
based on $^{13}$CO(3-2)  assuming optically thin emission (``$^{13}$CO Masses")
 differ by an order of magnitude. 
 There are many possible reasons for the differences. 
 For optically thin clouds, 
 D1, D2,  the  values of [CO]/[H$_{2}$] and the [$^{13}$CO]/[$^{12}$CO ]
 abundance ratios are sources of systemic uncertainty, 
 since they may differ significantly from Galactic values in this low metallicity galaxy. 
 For optically thick clouds, the 
 conversion factor cannot account for clouds that are unusually warm; or that have significant 
 stellar mass contained within the cloud, such as Clouds D1 and  D2; or for clouds that have
 significant CO-dark gas, although this is not likely to be a major mass contributor
 in these dense clouds
 \citep{2014A&A...561A.122L}.

 The CO(3-2) critical density is
 aligned with the gas densities expected for star-forming gas. 
 Even the highest estimates of cloud mass indicate that the
 gas masses of these clouds are fairly small. The largest cloud,
  D4, has an  estimated mass of 3.4$\times 10^5~\rm M_\odot$, a relatively low mass considering the $\sim 10^5~
\rm M_\odot$ star clusters that NGC 5253 has been forming. The current suite of clouds
does not seem capable of forming a massive super star cluster, unless mergers of clouds
take place, perhaps along accreting filaments, and if the star formation is extraordinarily efficient.

\section{Summary and Conclusions}

We present new ALMA images of $^{12}$CO(3--2) and $^{13}$CO(3--2) emission from the
nearby (3.8 Mpc) dwarf galaxy NGC 5253.  The spatial resolution is
$\sim$0.3\arcsec, 10 times smaller in diameter than the best previous maps.   
The images reveal dense ($\sim 20,000~\rm cm^{-3}$) molecular clumps $\sim$5~pc in diameter
with estimated masses of a few thousand $\rm M_\odot$. 

While atomic and ionized filaments are a feature of NGC 5253
on the largest scales \citep{2008AJ....135..527K},
 these observations are the first to detect clear, kinematically distinct filaments 
 in the molecular gas. 
The inner 200 pc of NGC 5253 holds four filaments, composed of 
small ($\sim$5~pc diameter) clouds. These filaments extend to $\sim$ 50-150 pc in
length. 
 The Streamer filament, detected previously in CO(1-0) and CO(2-1) in
 lower resolution images, 
 is aligned along the prominent dust lane to the east of the galaxy. The Streamer
  is  red-shifted with respect to the systemic 
  velocity, and thus appears to be 
falling into 
  the galaxy. Also redshifted, and foreground, thus
  apparently infalling, 
   is the weaker Southern filament to the south of the galaxy center.
Cloud F, a filament southwest of the center, is  
blue-shifted with respect to the galaxy but shows no apparent extinction,
and may be on the far side of the galaxy.
We propose that the CO filaments form the inner
position of HI filaments that are falling into the galaxy from
outside. This model particularly explains the Streamer along the
prominent dust lane (redshifted) and Cloud F (blueshifted, no
apparent foreground extinction). 
The clearest filaments have length-to-width aspect ratios of $\approx$20.  Dense
clumps in the filaments 
are likely to be overpressured relative to ambient gas, and probably gravitationally supported; they may
be likely 
sites of future star formation.

The central cloud surrounding the supernebula, Cloud D, previously detected
as a bright CO(3--2) source, breaks up into multiple clouds.
Some of these clouds are optically thin in CO(3-2), as indicated
by  high $^{12}$CO/$^{13}$CO flux ratios of $\sim 12$--40, higher than the
average cloud ratio of $\sim$6-8. 
Cloud D1 is coincident with the giant HII 
region and supernebula; its flux ratio of $\rm ^{12}CO/^{13}CO\sim$40 suggests that the
[$^{12}$CO]/[$^{13}$CO] ratio in NGC 5253 is $\ga$40.
Cloud D2 appears to contain a smaller young cluster.
We propose that Clouds D1 and
D2 are
most likely 
 optically thin in CO(3--2) because the molecular gas is hot, 
due to the proximity of  young and massive star clusters, although it 
is also possible that optical depth effects in the presence of high
temperature gradients could produce the high observed ratios. 
These clouds are likely to be 
 brighter than would be expected
based on their gas columns alone.

From radio
 continuum flux densities and lower J CO lines, 
we  derive average gas depletion timescales of 0.3 Gyr for the Streamer and 8 
Myr for Cloud D1 coincident with the supernebula.  
The masses we obtain for the individual clumps
from the ALMA observations, from assumption of
optically thin emission, for the thin clouds, and from $^{12}$CO(3--2) emission, for the thick
clouds, differ by about an order of magnitude. Uncertainties in the H$_2$ masses are
due to unknown [CO]/[H$_2$] for this low metallicity, high radiation environment, 
unknown CO(3--2)/CO(1--0) 
ratio, and uncertainty in the conversion factor applicability. However, even the highest
estimates of mass for the clumps, $\sim 10^4-10^5~\rm M_\odot$, are  small
compared to the masses of the star clusters that are currently forming. This suggests
that filament mergers and/or highly efficient star formation is taking place in NGC 5253.
Further high resolution
CO observations may clarify these issues and allow for improved masses. Even with these uncertainties,
it appears that the star formation in the central Cloud D1 associated with the supernebula
is extraordinarily efficient.

\begin{deluxetable*}{lcccccccc}
	\tabletypesize{\scriptsize}
	\tablecaption{Source, $^{12}$CO and $^{13}$CO Fluxes, Ratio, and Masses$^{**}$}%
	\tablewidth{0pt}%
	\tablenum{1}%
	\tablehead{
          \colhead{Source}&%
          \colhead{ICRS RA$^{a}$}&%
          \colhead{ICRS Dec$^{a}$}&%
          \colhead{$v_{sys}$$^{b}$}&%
          \colhead{$^{12}$CO(3-2)$^{c}$}&%
          \colhead{$^{13}$CO(3-2)$^{c}$}&%
          \colhead{$\frac{^{12}CO}{^{13}CO}$}&%
          \colhead{$^{12}$CO Mass$^{d}$}&%
          \colhead{$^{13}$CO Mass$^{e}$}\\%
 	  \colhead{}&%
          \colhead{13$^{h}$39$^{m}$}&%
          \colhead{-31\degr38\arcmin}&%
          \colhead{ \kms}&%
          \colhead{Jy~\kms~}&%
          \colhead{Jy~\kms~ }&%
          \colhead{ratio}&%
          \colhead{10$^{3}$ M$_{\odot}$}&%
          \colhead{10$^{3}$ M$_{\odot}$}%
          }

\startdata
D1         & 55.965  & 24.36 & 385  & 2.4  & 0.06$\pm$0.05      & 40$^{+30}_{-20}$   & 100 & 3.0 \\
D2            & 56.12    & 23.1   & 394  & 0.8  & 0.07$\pm$0.02      & 12$\pm$4     &  33  & 0.6 \\
D3            & 56.03    & 22.9   & 395  & 1.7  & 0.28$\pm$0.04      & 6$\pm$1     &    70  & 2.6  \\
D4            & 55.91    & 25.0   & 402  & 8.4  & 1.2$\pm$0.07      & 7$\pm$0.4    &    345    & 11  \\
D5            & 55.79    & 25.6   & 384  & 0.87  & 0.12$\pm$0.02      & 7$\pm$1     &     36     &  1.1 \\
D6            & 55.88    & 27.0   & 409  & 1.8  & 0.25$\pm$0.04      & 7$\pm$1     & 74  &   2.3 \\
Squiggle (D7) & 56.0     & 27     & 403  & 4.1  & 0.5$\pm$0.1     & 9$\pm$2     & 170 & 4.4  \\
Center Sum of D Clouds   & 55.95   & 25   & 385   & 20 & 2.4$\pm$0.1   & 8$\pm$0.5      &   823         & 24.7  \\
\hline
Streamer$^{f}$   & 56.5     & 33     & 430  & 8.79 & 0.86$\pm$0.2      & 10$\pm$2           & 360 & 8.1  \\
\hline
Cloud E             & 55.5     & 19     & 400  & 12.55  & 0.8$\pm$0.2         &  16$\pm$4     & 510   & 7.5 \\
\hline
Cloud F              & 55.8      & 31     & 368   & 6.82   & 0.9$\pm$0.1       & 8$\pm$1          & 280 & 8.2   \\
\hline
Southern Filament & 55.9 & 367    & 413  & 2.45  & 0.54$\pm$0.1         & 5$\pm$1      &  100           & 5.1  \\
\hline
Overall Sum         & 56.3   & 30     &  380    & 51    &    5.5
&    9                 &     2073      &  53.6 \\
\enddata
\label{tab:13co}

\tablenotetext{**}{All measurements are made from velocity-selected, primary beam corrected MOM0 integrated flux density maps.
  As $^{13}$CO emission is weak, $^{12}$CO was used to define velocity regions for each cloud, and all flux was summed for
  that velocity range.}
\tablenotetext{a}{Precision in coordinates depends on the compactness and  structure of the source.}
\tablenotetext{b}{$\pm$ 5 \kms}
\tablenotetext{c}{Uncertainties in $^{12}$CO flux density are  10\% from flux density calibration uncertainties 
unless otherwise stated. Uncertainties in $^{13}$CO flux density, which is mostly signal-to-noise limited,
 are computed from the individual channel rms times 
$\sqrt{N_{chan}N_{beam}}$. }
\tablenotetext{d}{Mass derived assuming optically thick emission, using a conversion factor of $\alpha_{CO}$=4.7$\times$10$^{20}$ cm$^{-2}$ (K kms$^{-1}$)$^{-1}$, and an estimated
CO(3--2)/CO(1--0) ratio of 7. This mass will overestimate
the cloud mass for optically thin, warm clouds and 
for clouds in which enclosed stellar mass is
significant. Mass includes He.}
\tablenotetext{e}{Mass derived assuming optically thin $^{13}$CO flux density and 20 K, a Galactic
value of [CO]/[H$_2$]=$8.5\times 10^{-5}$ and an isotopic ratio of 40, except for Cloud D1, where T=300 K is assumed. 
The Rayleigh-Jeans approximation is not used. 
This mass will underestimate the true cloud mass for optically thick emission or 
if the clouds 
are warmer than 20 K.  Mass includes He.}
\tablenotetext{f} {All integrated line intensities, particularly
  $^{12}$CO, are subject to the missing short spacing data, and
  therefore 
  the inability of this 
image to reconstruct structures $>$4\arcsec\ in size 
This will not affect small clouds such as the Cloud D individual clumps. 
The Streamer, however, 
 appears to be
missing flux on the whole. See text. }
\end{deluxetable*}

\acknowledgments

This paper makes use of the following ALMA data: ADS/JAO.ALMA\# 2012.1.00105.S.
ALMA is a partnership of ESO (representing its member states), NSF
(USA) and NINS (Japan), together with NRC (Canada), MOST and ASIAA
(Taiwan), and KASI (Republic of Korea), in cooperation with the
Republic of Chile. The Joint ALMA Observatory is operated by ESO,
AUI/NRAO and NAOJ. The National Radio Astronomy Observatory (NRAO) is
a facility of the National Science Foundation operated under
cooperative agreement by Associated Universities, Inc.  Support for
this work was provided by the NSF through award GSSP SOSPA2-016 from
the NRAO to SMC and grant AST 1515570 to JLT, and by the UCLA Academic
Senate through a COR seed grant. The authors wish to thank the
W. M. Goss, P. T. P. Ho, and the anonymous referee for their helpful
comments. 

{\it Facilities:} \facility{ALMA}.

\end{document}